\documentclass[%
 reprint,
nofootinbib,
 amsmath,amssymb,
 aps,
]{revtex4-1}

\usepackage[dvips]{color}
\usepackage{graphicx}
\usepackage{dcolumn}
\usepackage{bm}

\newcommand{\tr}{{\rm tr}}

\newcommand{\mev}{{\rm MeV}}
\newcommand{\gev}{{\rm GeV}}

\newcommand{\lmd}{\lambda}
\newcommand{\TeV}{\mbox{TeV}}
\newcommand{\GeV}{\mbox{GeV}}

\newcommand{\tz}{{T^0}}
\newcommand{\tp}{{T^+}}
\newcommand{\tpp}{{T^{++}}}
\newcommand{\tzr}{{T_r^0}}
\newcommand{\tzi}{{T_i^0}}
\begin{document}

\preprint{APS/123-QED}

\title{Dark Matter in Inert Triplet Models}

\author{Takeshi~Araki$^1$}
\email{araki@ihep.ac.cn}

\author{C.~Q.~Geng$^{2,3}$}
\email{geng@phys.nthu.edu.tw} 

\author{Keiko~I.~Nagao$^2$}
\email{nagao@phys.nthu.edu.tw}
\affiliation{$^1$Institute of High Energy Physics, Chinese Academy of Sciences, Beijing 100049, China\\
$^2$Department of Physics, National Tsing Hua University, Hsinchu, Taiwan 300\\
$^{3}$National Center for Theoretical Sciences, Hsinchu, Taiwan 300}


\begin{abstract}
We study the inert triplet models, in which the standard model (SM) is extended to have a new $SU(2)_L$ triplet scalar (Y=0 or 2) with an $Z_2$ symmetry.
We show that the neutral component of the triplet can be a good dark matter candidate.
In particular, for the hypercharge Y=0 triplet model, the WMAP data favors the region where the dark matter mass is around $5.5\ \TeV$, 
which is also consistent with the direct detection experiments.
 In contrast, for the Y=2 model,  although dark matter with its mass around $2.8~\TeV$ is allowed by WMAP, 
 it is excluded by the direct detection experiments
because the spin-independent cross section is enhanced by the $Z$ mediated tree-level scattering process.
\end{abstract}


\maketitle


\section{Introduction}
  From the Wilkinson Microwave Anisotropy Probe (WMAP) observation, the relic abundance of the cold dark matter in the universe is determined to be \cite{Komatsu:2010fb}
\begin{eqnarray}
\Omega_{CDM} h^2=0.1123 \pm 0.0035,
\label{eq:WMAP_relic}
\end{eqnarray}
where $h = 0.710 \pm 0.025$ is the scaled current Hubble parameter in units of $100\,\mathrm{km\, sec^{-1}\, Mpc^{-1}}$.
Since the standard model (SM) cannot accommodate 
dark matter,
new physics is expected.
In this paper, we study dark matter in a model  containing an $SU(2)_L$ triplet scalar with the hypercharge Y=0 or 2
under $U(1)_Y$, which is  clearly  one of the minimal  extensions of the SM.
In the model, the triplet is odd under an $Z_2$ symmetry so that it neither directly couples to the SM fermions nor develops a
vacuum expectation value (VEV). We will refer to the model as
the inert triplet model (ITM).
Since  the neutral component of the triplet scalar can be the lightest one and stable in both Y=0 and 2 cases,
it is a good dark matter candidate.

It can be shown that there are three and five new parameters in the Y=0 and 2 ITMs, which are the same as those in 
the inert singlet and doublet models \cite{Silveira:1985rk,Ma:2006km, Cirelli:2009uv, Hambye:2009pw}\footnote{In \cite{Hambye:2009pw},  the ITM was also mentioned.}, respectively.
Clearly, the Y=0 ITM is one of the minimal inert models.

Besides the relic abundance, the direct searches of dark matter also provide constraints on 
new physics models.
For the spin-independent (SI) cross section, 
one has that
\begin{eqnarray}
\sigma_{SI}\lesssim 5\times 10^{-44}- 10^{-42} \mathrm{cm^2}
\label{eq:DD_constraint}
\end{eqnarray}
for the range of the Weakly Interacting Massive Particle (WIMP) mass smaller than $10^3~\GeV$~\cite{Angle:2007uj}.
Note that if the dark matter mass is larger,
the constraint in Eq.~(\ref{eq:DD_constraint}) 
will be relaxed.

The  paper is organized as follows. 
In Sec. \ref{sec:Y=0}, we introduce the ITM of the Y=0 case and discuss the relic abundance 
as well as the direct detection of dark matter. 
We extend our study to the Y=2 ITM in Sec.~\ref{sec:DMY=2}.
We conclude in Sec.~\ref{sec:conclusion}.
\section{Dark Matter in Y=0 ITM}
\label{sec:Y=0}
\subsection{Basic framework}
In addition to the SM particles, we introduce an $SU(2)_L$ 
triplet scalar with Y=0 and impose an $Z_2$ symmetry 
in which the triplet is assigned to be odd and the others even.
Furthermore, we assume that the triplet scalar does not 
develop the VEV to keep the $Z_2$ symmetry unbroken.
The relevant Lagrangian
is given by
\begin{eqnarray}
{\cal L}&=&|D_\mu H|^2 + \tr|D_\mu T|^2 - V(H,T), 
\nonumber\\
V(H,T)&=&
    m^2 H^\dag H + M^2 \tr[T^2] + \lambda_1 |H^\dag H|^2 
   \nonumber\\
  &&+ \lambda_2 \left(\tr[T^2]\right)^2+ \lambda_3 H^\dag H\ \tr[T^2] \label{eq:3vertex},
\end{eqnarray}
where $D_\mu$ is the covariant derivative and the  doublet $H$ and  triplet 
 $T$ scalars are defined as
\begin{eqnarray}
H=\left(\begin{array}{c} 
   \phi^+ \\ 
   \frac{1}{\sqrt{2}}(h +i\eta)
  \end{array}\right),\ \ \ 
T=\left(\begin{array}{cc}
   \frac{1}{\sqrt{2}}T^0 & -T^{+} \\
   -T^- & -\frac{1}{\sqrt{2}}\tz
  \end{array}\right),
\end{eqnarray}
 with 
 $\langle h\rangle=v=246 ~\gev$ and 
$\langle \tz\rangle=0$, respectively. 
In order to assure the stability of the potential, 
we require the conditions  
\begin{eqnarray}
 \lambda_1,\ \lambda_2 > 0 ,~~\,2\sqrt{\lambda_1 \lambda_2} > |\lambda_3|\ \ {\rm for\ negative}\ \lambda_3.
\end{eqnarray}
The potential in Eq. (\ref{eq:3vertex}) becomes a local minimum if and only if
\begin{eqnarray}
m^2 <0,\ \ \ 2M^2 + \lambda_3 v^2 > 0\ ,
\end{eqnarray}
where $v^2 = - m^2/\lambda_1$.
After $h$ acquires the VEV, the scalars gain 
the following masses:
\begin{eqnarray}
m_h^2 = 2\lambda_1 v^2\ , 
m_{\tz}^2 = m_{T^\pm}^2 = M^2 + \frac{1}{2}\lambda_3 v^2.
\end{eqnarray}
Note that $\eta$ and $\phi^{\pm}$ are the massless Nambu-Goldstone 
bosons eaten by the SM gauge fields.
Although the masses of $\tz$ and $T^\pm$ are degenerate at 
the tree level, a small mass splitting
\begin{eqnarray}
m_{T^\pm}=m_{\tz} + (166\ \mev) \label{eq:gap}
\end{eqnarray}
will appear  once the
radiative corrections \cite{Cirelli:2009uv} 
are taken into account.
Hence, $T^0$ turns out to be the lightest component of 
the triplet scalar and moreover,  it is stable due to $Z_2$.

\subsection{Oblique parameters and Higgs boson mass}
Since the triplet scalar is added to the SM, one may 
think that it affects the so-called oblique 
(S and T) parameters.
In general, however, an Y=0 triplet 
has no
contribution  to 
the S parameter, while the contribution to the T parameter is 
also vanishing in the limit of $m_{T^0} = m_{T^\pm}$ \cite{oblq}.
Even if we consider
the mass splitting in 
Eq. (\ref{eq:gap}), 
its effect is negligibly small. 
Therefore, the constraint on the Higgs boson mass ($m_h$) from 
the precision electroweak measurements is the 
same as that in the SM.
In our calculation, we restrict $m_h$
to be within the range of
\begin{eqnarray}
114\ \gev<m_h<185\ \gev
\end{eqnarray}
as estimated in Ref. \cite{LEP} 
with the excluded region of $158 \sim 175\ \gev$ 
reported by the Tevatron \cite{tev}.

\subsection{$W$ and $Z$ decay widths and DM mass}
Since the decay widths of  $Z$ and $W$  precisely 
measured by the LEP and Tevatron,
 agree well with the SM predictions, 
the new decay processes $W^\pm \rightarrow T^\mp T^0$
and $Z\rightarrow T^\pm T^\mp$ must be strongly suppressed.
To this end, we impose the following condition: 
\begin{eqnarray}
m_\tz -(166\ \mev)>m_Z/2 \label{eq:masssplit}.
\end{eqnarray}

\subsection{Relic Abundance}
We now examine  the thermal relic abundance of $\tz$.
The evolution of the number density of $\tz$ is obtained by solving the 
Boltzmann equation
\begin{eqnarray}
\frac{dn_\tz}{dt}+3Hn_\tz=-\langle \sigma v_\tz\rangle (n_\tz^2-n_{\tz,\,eq}),
\label{eq:Boltzmann}
\end{eqnarray}
where $H$ is
the Hubble parameter, $v_\tz$ stands for relative
velocity of\, $T^0$, 
$\langle \cdots \rangle$ represents the thermal average of a function 
in brackets, and $n_\tz$, $n_{\tz,\,eq}$ and $\sigma$
are the number density, the number density  in thermal equilibrium
 and the total annihilation cross section of $\tz$, respectively.
In the model, since the mass splitting between dark matter ($T^0$) and charged components ($T^\pm$) 
is much smaller than their masses,
 the coannihilation effects of $\tz\,T^\pm$ and $T^\pm \,T^\mp$ should
be included in $\sigma$~\cite{Griest:1990kh}.

\begin{figure}[htbp]
\begin{center}
\includegraphics[width=5cm,clip]{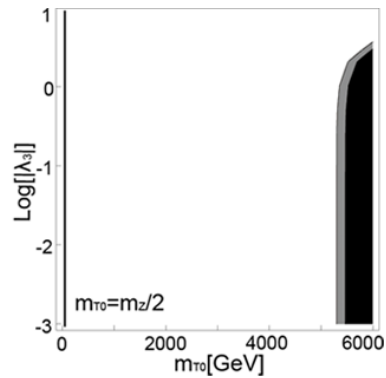}
\caption{Relic abundance for $m_h=120~\GeV$ with the vertical axis of $\mathrm{Log}[\lmd_{3}]$, where black, gray, and white regions show the parameter regions larger than, agreed with, and smaller than the WMAP constraint, respectively, while
the straight line indicates the LEP bound for $\tz$.}
\label{fig:relic_abundance}
\end{center}
\end{figure}
In Fig. \ref{fig:relic_abundance}, we show the relic abundance of \,$\tz$, where we 
have used \textit{micrOMEGAS\,2.4}\cite{Belanger:2010gh} to scan the parameter $\lmd_3$ from $10^{-3}$ to $10$. 
For small couplings, i.e. $\lmd_3 \lesssim 1$, the dark matter annihilation is governed by the
weak interaction. So the annihilation cross section does not decrease so much.
In this case, the main (co)annihilation modes are $\tz\,T^\pm\to \gamma W^\pm$,
$\tz\,\tz \to W^+W^-$ and $T^+\,T^-\to W^+W^-$.
On the other hand, in the large coupling region (i.e. $\lmd_3 \gtrsim 1$), the main annihilation modes are
$\tz \,\tz \to t \bar{t}$ and $\tz \,\tz \to hh$. 
Although some of those annihilation are mediated by the Higgs $h$, the relic abundance 
 is not subject to $m_h$
so much
as long as we take $m_h=114\sim 185~\GeV$.
Since the trilinear coupling of $h$ involves only $\lambda_3$ (see Eq. (\ref{eq:3vertex})), 
the cross sections are enhanced.
Note that the relic abundance depends on only $\lmd_{3}$, whereas
both $\lmd_1$ and $\lmd_2$ are irrelevant to the annihilation interactions.
From the figure, we find that for $5.4\ \TeV \lesssim m_\tz \lesssim 6\ \TeV$, the relic abundance agrees with 
the WMAP data in Eq.~(\ref{eq:WMAP_relic}).
We remark that $m_\tz \lesssim 5000\ \GeV$ is allowed if there are some other dark matter sources.

\subsection{Direct Detection}

\begin{figure}[htbp]
\begin{center}
\includegraphics[width=5cm,angle=0,clip]{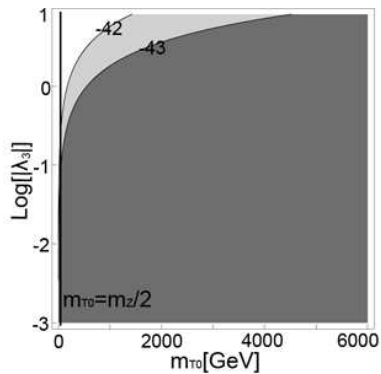}
\caption{Spin-independent scattering cross section between the dark matter $\tz$ and nucleus particles, where numbers on lines represent the cross sections in $cm^{2}$ unit, while light gray and dark gray regions are allowed by direct searches of dark matter. }
\label{fig:direct_detection}
\end{center}
\end{figure}
The SI cross section of the Y=0 ITM is shown in Fig. \ref{fig:direct_detection}.
>From the figure, we can see that in most of the region, the model escapes the constraint from the direct search.
We note that 
 the contribution of the SI cross section is insensitive to $m_h$ as long as $114~\GeV\leq m_h \leq185~\GeV$ even though it comes from
  the $\tz$-quark and $\tz$-gluon collisions through the $\tz-\tz-h$ coupling.
Since the $\tz$-quark (u,d) scattering has a small cross section due to  the small Yukawa couplings,
while $\tz$-gluon scattering occurs only in loop level,
the SI cross section is clearly suppressed.

\section{Dark Matter in Y=2 ITM}
\label{sec:DMY=2}
\subsection{Basic framework}
In the model with  the inert triplet scalar of Y=2,
 the $Z_2$ invariant scalar potential is 
given by
\begin{eqnarray}
V (H,T)&=& 
    m^2 H^\dag H + M^2 \tr[T^\dag T] + \lambda_1 |H^\dag H|^2 \nonumber\\
  && + \lambda_2 \tr[T^\dag T T^\dag T] + \lambda_3 \left( \tr[T^\dag T] \right)^2
  \nonumber \\
&&+ \lambda_4 H^\dag H\ \tr[T^\dag T] + \lambda_5 H^\dag T T^\dag H\ ,
\label{eq:Higgs_Y2}
\end{eqnarray}
where
\begin{eqnarray}
T=\left(\begin{array}{cc}
   \frac{1}{\sqrt{2}}T^+ & T^{++} \\
   \tzr + i\tzi & -\frac{1}{\sqrt{2}}T^+
  \end{array}\right).
\end{eqnarray}
The masses of the scalars are calculated as 
\begin{eqnarray}
&& m_h^2 = 2\lambda_1 v^2\ , 
\nonumber\\
&& m_\tzr^2 = m_\tzi^2 = M^2 + \frac{1}{2}(\lambda_4 + \lambda_5)v^2\ ,
\nonumber \\
&& m_{T^{\pm}}^2 = M^2 + \frac{1}{2}\left(\lambda_4 + \frac{\lambda_5}{2} \right)v^2 
                 = m_{\tzr(\tzi)}^2 - \frac{\lambda_5}{4}v^2\ ,
                 \nonumber\\
&& m_{T^{\pm\pm}}^2 = M^2 + \frac{1}{2}\lambda_4 v^2 = m_{\tzr(\tzi)}^2 - \frac{\lambda_5}{2} v^2\ .
\end{eqnarray}
We note that, in order to make $\tzr$ and $\tzi$ to be the 
lightest $Z_2$-odd particles, we take $\lambda_5 < 0$ afterward.

\subsection{Relic abundance}

\begin{figure}
\includegraphics[width=50mm]{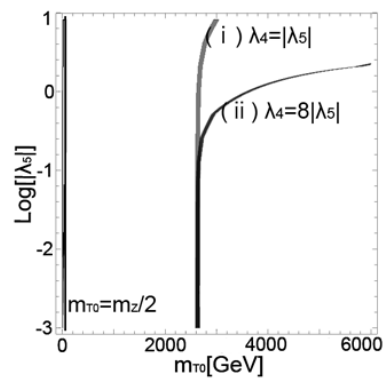}
    \caption{Relic abundance of Y=2 ITM for $m_h=120~\gev$,
    where gray regions are the parameter regions agreed with WMAP and  the right (left) handed sides of the gray regions 
    correspond to the regions larger (smaller) than the observation. }
    \label{fig:relic_abundance2}
\end{figure}

The total relic abundance of $\tzr$ and $\tzi$ is shown\footnote{Here, we again use {\it micrOMEGAS} and assume that  the interactions of $\tzr$ and $\tzi$ are the same, which simplifies our calculation to the relic abundance of $\tzr$ only. 
It is justified since the interactions of $\tzr$ and $\tzi$ are almost the same.} in Fig. \ref{fig:relic_abundance2}.
Note that the masses of $T^\pm$ and $T^{\pm\pm}$ are automatically fixed if $m_{\tzr}$ and $\lmd_5$ are known.
It is easy to see
 that the  relic abundance tends to be large compared to that in the  Y=0 case. 
Moreover, the mass splitting among the triplet scalars is not so small unless the absolute value of $|\lmd_5|$ is enormously small.
Since coannihilations of the triplet scalars are not so effective,  the relic abundance gets enhanced.
However, in the small $\lmd_5$ region (i.e., $|\lmd_5|\lesssim 1$), the masses of $\tzr$, $\tp$ and $\tpp$ are still degenerate.
As the result, the coannihilation involving $\tp$ and $\tpp$ 
(e.g., $T^+\, T^{--} \to \gamma\, W^-$ and $T^+T^-\to W^+ \,W^-$) 
is active.
In the large $|\lmd_5|$ region, as the mass degeneracy of the triplet components is lifted, the coannihilation effect becomes weaker,
which enhances the relic abundance.
However, the annihilation cross section becomes large due to the large couplings of  $\lmd_4$ and $|\lmd_5|$, which suppresses the relic abundance more effective than the coannihilation effect.

\begin{figure}[t]
\begin{center}
\includegraphics[width=7.cm,clip]{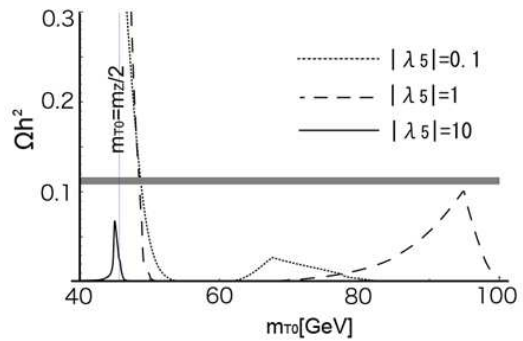}
\caption{Relic abundance in the Y=2 ITM for 
the small $m_{T^0_r}$ region with $m_h=120~\GeV$, 
where the light gray region is  allowed by WMAP.
}
\label{fig:relic_abundance100}
\end{center}
\end{figure}
In the region with $m_\tzr \lesssim 100$~GeV, the relic abundance drastically changes due to the resonance effect 
as well as the opening of new annihilation final states. We show the relic abundance in the small mass region in Fig.~\ref{fig:relic_abundance100}.
In the figure, we have  fixed $\lmd_4=|\lmd_5|/8$. 
For $|\lmd_5|=10$, the relic abundance tends to be small due to the large coupling.
The main annihilation mode is $\tzr \tzr (\tzi \tzi)\to b\bar{b}$, while $\tzr \tzr (\tzi \tzi)\to Z h$ is also effective if $m_\tzr$  is large.
For $|\lmd_5|=1$, we find that the relic abundance suddenly decreases at $m_\tzr\sim 60$~GeV and $m_\tzr\sim 100$~GeV
due to $\tzr \tzr (\tzi \tzi) \to h \to b\bar{b}$ and $\tzr \tzr (\tzi \tzi) \to hh$, respectively.
 In contrast, for $|\lmd_5|=0.1$, the main interaction mode at the large $m_\tzr$ region is 
 $\tzr\tzr (\tzi \tzi)\to W^+ W^-$
  since 
  the gauge interaction is more effective than the Higgs interaction. 
 In this case,   the relic abundance becomes much  smaller.
 For a smaller $|\lmd_5|$ (i.e. $|\lambda_5|=0.01$), the figure is similar to 
 that for the case of $|\lmd_5|=10$ as 
the relic abundance is small due to 
the large coannihilation cross section of  $T^-T^{++} \to \gamma W^+$, $\tzr (\tzi) T^+\to \gamma W^+$ and $\tzr (\tzi) T^+\to Z W^+$.
We note that for $\lmd_4=|\lmd_5|$,  the figures are similar to those with $|\lmd_5|=\lmd_4/8=0.1$ and $0.01$ since the main interactions enhanced by the large Higgs coupling are proportional to $(\lmd_4+\lmd_5)$.

In the region with $|\lmd_4|/|\lmd_5|\neq 1$, where $\tzr \,\tzr (\tzi \tzi)\to h\,h$  is most effective,
the relic abundance is reduced.
In the case of $|\lmd_4|/|\lmd_5|= 1$, the tri-Higgs couplings proportional to $(\lmd_4+\lmd_5)$ are canceled to be $0$.
Since the effective couplings of the Higgs bosons are very weak, 
the relic abundance is determined by gauge interactions.

We comment on the direct detection of the Y=2 case.
Unlike Y=0, there are three scattering processes in the Y=2 model. 
Two of them are the same as those in the Y=0 case,
while the other one is the $\tz$-quark scattering through the gauge coupling of $\tz$ to $Z$ as shown 
explicitly in Appendix.
The latter has a larger cross section due to  the gauge coupling.
Because of this large cross section, almost  all region is excluded by the direct detection constraint in Eq.~(\ref{eq:DD_constraint})\footnote{The DM-DM-Z coupling tends to make the SI cross section beyond the constraint of the direct search, which is consistent with the result in Ref.~\cite{Chun:2009mh}}. 
In particular, we have checked that in all of the regions allowed by LEP experiments, 
the SI cross section is larger than about $10^{-37} \mathrm{cm^2}$.
Therefore, even if the ratio of $\lmd_4$ and $\lmd_5$ (i.e., the coupling of DM-gluon scattering) is changed,
the cross section is still larger 
than the constraint from the direct detection. 

\section{Conclusion}
\label{sec:conclusion}
We have studied dark matter in the two inert triplet models. In the Y=0  model, we 
have shown that the favored region by the WMAP result is around $m_\tz \sim 5.5~\TeV$ based on the relic abundance. 
On the other hand, since $\tz$  scatters quarks  only for the small Yukawa couplings
as it does not couple to $Z$ at the Lagrangian level, while the $\tz$-gluon scattering occurs at loop level,
dark matter ($\tz$) in most of the regions, including that favored by WMAP, is allowed
from the direct detection.
For the Y=2 case, $m_\tz\sim 2.8\, \TeV$ is preferred in terms of the relic abundance of $\tz$. However, 
since the $\tz$-quark scattering is allowed at tree level due to the $\tz - Z$ coupling, which enhances the scattering cross section,
most of the regions is excluded by the direct detection.\\
\\
{\bf Acknowledgement}
We are grateful to G.~B$\mathrm{\acute{e}}$langer and A.~Pukhov for their kind help for {\it micrOMEGAs}.
The work of T.A. was supported in part by the National
Natural Science Foundation of China under Grant No. 10425522 and No.
10875131. C.Q.G. and K.I.N were partially supported  by the National Science
Council of Taiwan under Grant No. NSC-98-2112-M-007-008-MY3 and the National
Tsing Hua University under the Boost Program No. 97N2309F1.

\section{Appendix: Interactions}
\label{sec:app}
We expand Eqs. (\ref{eq:3vertex}) and (\ref{eq:Higgs_Y2}) 
in terms of the component fields to show specific scalar 
and gauge interactions of the triplet.
In the followings, we will use the following definitions 
of the gauge fields:
\begin{eqnarray}
&&D_\mu = \partial_\mu -ig\left[ W_\mu^a\frac{\sigma^a}{2},~~~\right] -ig^\prime\frac{Y}{2}B_\mu\\
&&W_\mu^{\pm} = \frac{1}{\sqrt{2}} (W_\mu^1 \mp iW_\mu^2), \\
&&Z_\mu = c_w W_\mu^3 - s_w B_\mu ,~~
A_\mu = s_w W_\mu^3 + c_w B_\mu ,
\end{eqnarray}
where $\sigma^{a=1\cdots 3}$ are the Pauli matrices, 
$s_w(c_w)=\sin\theta_w(\cos\theta_w)$, and 
$\theta_w$ is the weak mixing angle.
\subsection{Interactions in Y=0 ITM}
The scalar interactions:
\begin{itemize}
\item
$m^2 H^\dag H = 
m^2 \left[ 
|\phi^+|^2 + \frac{1}{2}(h^2 + \eta^2) 
\right],$
\item
$M^2 \tr[T^2] = 
M^2 \left[ 2|T^+|^2 + (\tz)^2 \right],$
\item
$\lambda_1 |H^\dag H|^2 = 
\lambda_1 \left[ 
|\phi^+|^2  + \frac{1}{2}(h^2 + \eta^2) 
\right]^2,$
\item
$\lambda_2 \left(\tr[T^2]\right)^2 = 
\lambda_2 \left[ 2|T^+|^2 + (\tz)^2 \right]^2,$
\item
$\lambda_3 H^\dag H~\tr[T^2]$ \vspace{1mm}\\
$=\left[ 
|\phi^+|^2 + \frac{1}{2}(h^2 + \eta^2) 
\right]
\left[ 2|T^+|^2 + (\tz)^2 \right].$
\end{itemize}
The three-point gauge interactions:
\begin{itemize}
\item 
$2ig \left[ (\partial^\mu T^+)W_\mu^- \tz +(\partial^\mu T^0)W_\mu^+ T^- \right] + h.c.~,$
\item
$2ig (\partial^\mu T^+) (c_w Z_\mu + s_w A_\mu)T^- + h.c.~,$
\end{itemize}
and four-point gauge interactions:
\begin{itemize}
\item
$g^2\left[ |W_\mu^- T^+ - W_\mu^+ T^-|^2 + 2|W_\mu^+ \tz|^2 \right],$
\item
$2g^2 (c_w Z_\mu + s_w A_\mu)^2 |T^+|^2 ,$
\item
$2g^2 (W_\mu^+ \tz) (c_w Z_\mu + s_w A_\mu) T^- + h.c.~.$
\end{itemize}
Notice $\tz$ does not couple to Z boson in this model.

\subsection{Interactions in Y=2 ITM}
The scalar interactions:
\begin{itemize}
\item
$M^2\tr[T^\dag T] = 
M^2 \left[ |T^{++}|^2 + |T^+|^2 + \tzr^2 + \tzi^2 \right],$
\item
$\lambda_2 \tr[T^\dag T T^\dag T]$ \\ 
$=\lambda_2 \left[ \frac{1}{2}|T^+|^4 + |T^{++}|^4 + (\tzr^2 + \tzi^2)^2 \right.$ \\ 
$~~~~~~ +2|T^+|^2(\tzr^2 + \tzi^2)+ 2|T^+|^2|T^{++}|^2 $ \vspace{2mm} \\
$~~~~~~ -\left. \left\{T^- T^{++} T^- (\tzr+i\tzi)+h.c.\right\} \right] ,$
\item
$\lambda_3 \left( \tr[T^\dag T] \right)^2$ \\
$= \lambda_3 \left[ |T^{++}|^2 + |T^+|^2 + \tzr^2 + \tzi^2 \right]^2 ,$ 
\item
$\lambda_4 H^\dag H\ \tr[T^\dag T]$ \vspace{1mm} \\ 
$= \lambda_4 \left[ |\phi^+|^2 + \frac{1}{2}(h^2 + \eta^2) \right]$ \\
$~~~~~~~~~~\times \left[ |T^{++}|^2 + |T^+|^2 + \tzr^2 + \tzi^2 \right],$
\item
$\lambda_5 H^\dag T T^\dag H$ \\
$= \lambda_5 \left[ \frac{1}{2}|\phi^+|^2|T^+|^2 + \frac{1}{2}(h^2 + \eta^2)(\tzr^2 + \tzi^2)\right.$ \vspace{1mm} \\
$~~~~~~~ + |\phi^+|^2|T^{++}|^2 + \frac{1}{4}(h^2 + \eta^2)|T^+|^2$ \vspace{1mm} \\
$~~~~~~~ +\frac{1}{2}\left\{\phi^- T^+ (\tzr - i\tzi)(h + i\eta)\right.$ \vspace{1mm} \\
$~~~~~~~~~~~~~~~~~~~~\left.\left. - \phi^- T^{++} T^- (h + i\eta) + h.c.\right\}\right].$
\end{itemize}
The three point gauge interactions:
\begin{itemize}
\item
$ig\left[ (\partial^\mu T^+)\left(W_\mu^- (\tzr-i\tzi)- W_\mu^+ T^{--}\right)\right.$ \vspace{1mm} \\
$~~~~~~~~~~~~~~~~~~~ -(\partial^\mu T^{++})W_\mu^- T^- $ \vspace{1mm} \\
$~~~~~~~~~~~~~~~~~~~ \left. + \left(\partial^\mu \tzr+i\partial^\mu\tzi\right)W_\mu^+ T^-\right]+h.c.~,$
\item
$i\left[ (gc_w - g^\prime s_w)Z_\mu + 2gs_w A_\mu \right](\partial^\mu T^{++})T^{--} +h.c.~,$
\item
$\frac{4m_Z}{v}\left[ -(\partial^\mu \tzr)\tzi + (\partial^\mu \tzi)\tzr \right] Z_\mu ~,$
\item
$ig^\prime \left(-s_w Z_\mu + c_w A_\mu \right)(\partial^\mu T^+)T^- +h.c.~,$
\end{itemize}
and four-point gauge interactions:
\begin{itemize}
\item
$g^2\left[ |W_\mu^+|^2 (\tzr^2 + \tzi^2) \right.$ \\
$~~~~~~~~~~~~~~~\left. + |W_\mu^+|^2 |T^{++}|^2 + 2|W_\mu^+|^2|T^+|^2\right],$
\item
$\frac{4m_Z^2}{v^2}\ Z^\mu Z_\mu (\tzr^2 + \tzi^2)~,$
\item
$\left[ (g^2-g^{\prime 2})(c_w^2 - s_w^2)Z_\mu Z^\mu + 4g^2 s_w^2 A_\mu A^\mu \right. $ \vspace{1mm}\\
$~~~~~~~~~~~~~~~~~~\left.+ 4gg^\prime (c_w^2 - s_w^2)Z_\mu A^\mu\right] |T^{++}|^2 ,$
\item
$\left[ (-g^2 c_w + 2gg^\prime s_w)Z^\mu -3 g^2 s_w A^\mu\right]$ \vspace{1mm}\\
$~~~~~~~~~~~~~~~~~~~~~~~~~~~~~~\times T^{++} W_\mu^- T^- + h.c.~,$
\item
$\left[ (-g^2 c_w - 2gg^\prime s_w)Z^\mu + g^2 s_w A^\mu\ \right]$ \vspace{1mm}\\
$~~~~~~~~~~~~~~~~~~~~~~\times(\tzr+i\tzi) W_\mu^+ T^- + h.c.~,$
\item
$g^{\prime 2}(s_w^2 Z^\mu Z_\mu + c_w^2 A^\mu A_\mu - 2s_w c_w Z_\mu A^\mu)|T^+|^2 .$
\end{itemize}
Unlike Y=0 case, both $\tzr$ and $\tzi$ couple to Z boson.


\end{document}